\documentclass[conference]{IEEEtran}
\IEEEoverridecommandlockouts
\usepackage{cite}
\usepackage{amsmath,amssymb,amsfonts}
\usepackage{graphicx, enumitem}
\usepackage{textcomp}
\usepackage[ruled,vlined,linesnumbered]{algorithm2e}
\usepackage[table,dvipsnames]{xcolor}
\usepackage{tikz}
\usepackage{bm}
\usepackage{makecell}
\usepackage{booktabs}
\usepackage{multirow}
\usepackage{url}
\usepackage{subfigure}
\usepackage{tikz}
\usepackage[none]{hyphenat}
\SetKwIF{If}{ElseIf}{Else}{if}{}{else if}{else}{end if}%

\newcolumntype{L}[1]{>{\raggedright\let\newline\\\arraybackslash\hspace{0pt}}m{#1}}
\newcolumntype{C}[1]{>{\centering\let\newline\\\arraybackslash\hspace{0pt}}m{#1}}
\newcolumntype{R}[1]{>{\raggedleft\let\newline\\\arraybackslash\hspace{0pt}}m{#1}} 

\def\BibTeX{{\rm B\kern-.05em{\sc i\kern-.025em b}\kern-.08em
    T\kern-.1667em\lower.7ex\hbox{E}\kern-.125emX}}

\begin{document}
	
\title{Analyzing the behaviour of D'WAVE quantum annealer: fine-tuning parameterization and tests with restrictive Hamiltonian formulations\\
	{}
	\thanks{This work was supported by the Basque Government through HAZITEK program (BIZIBIKE project, ZE-2021/00031), and ELKARTEK program (QUANTEK project, KK-2021/00070), and by the Spanish CDTI through Misiones Ciencia e Innovación Program (CUCO) under Grant MIG-20211005.}
}

\author{
	\IEEEauthorblockN{Esther Villar-Rodriguez\IEEEauthorrefmark{2},
	    Eneko Osaba\IEEEauthorrefmark{2}\IEEEauthorrefmark{1}, and
		Izaskun Oregi\IEEEauthorrefmark{2}}
	\IEEEauthorblockA{\IEEEauthorrefmark{2}TECNALIA, Basque Research and Technology Alliance (BRTA), 48160 Derio, Bizkaia, Spain}
	\IEEEauthorblockA{\IEEEauthorrefmark{1}Corresponding author. Email: eneko.osaba@tecnalia.com}}
\maketitle

\IEEEpubidadjcol	

\begin{abstract}
Despite being considered as the next frontier in computation, Quantum Computing is still in an early stage of development. Indeed, current commercial quantum computers suffer from some critical restraints, such as noisy processes and a limited amount of qubits, among others, that affect the performance of quantum algorithms. Despite these limitations, researchers have devoted much effort to propose different frameworks for efficiently using these Noisy Intermediate-Scale Quantum (NISQ) devices. One of these procedures is D'WAVE Systems' quantum-annealer, which can be use to solve optimization problems by translating them into an energy minimization problem. In this context, this work is focused on providing useful insights and information into the behaviour of the quantum-annealer when addressing real-world combinatorial optimization problems. Our main motivation with this study is to open some quantum computing frontiers to non-expert stakeholders. To this end, we perform an extensive experimentation, in the form of a parameter sensitive analysis. This experimentation has been conducted using the Traveling Salesman Problem as benchmarking problem, and adopting two QUBOs: state-of-the-art and a heuristically generated. Our analysis has been performed on a single 7-noded instance, and it is based on more than 200 different parameter configurations, comprising more than 3700 unitary runs and 7 million of quantum reads. Thanks to this study, findings related to the energy distribution and most appropriate parameter settings have been obtained. Finally, an additional study has been performed, aiming to determine the efficiency of the heuristically built QUBO in further TSP instances.
\end{abstract}


\section{Introduction}
\label{sec:Introduction}

Researchers and practitioners have envisioned Quantum Computing (QC, \cite{steane1998quantum}) as the next frontier in computation. High hopes have been pinned recently on this paradigm, which has been already adopted for solving a wide variety of machine learning and optimization problems \cite{li2020quantum}. This paper is focused on the second of these categories, optimization, for which QC provides revolutionary mechanisms for facing complex problems with a remarkable advantage \cite{ajagekar2020quantum}. 

Today, two QC architectures coexist, gate-model and annealing-based quantum computers, which have been employed in the last few years for solving artificial intelligence problems. Nevertheless, QC is still in a budding stage of development, and current available QC computers (regardless of their architectures) suffer from some hardware limitations, such as noise and a restricted number of qubits, that critically affect the performance of quantum algorithms \cite{fellous2021limitations}.

This manuscript is focused on the quantum-annealers type, concretely on the one provided by D'WAVE Systems\footnote{https://www.dwavesys.com/}, the most used QC commercial hardware on the current literature. As previously pointed in the literature, quantum annealers are highly sensitive to the fine-tuning of involved parameters. Several published papers have spotlighed this issue, such as \cite{mehta2019quantum}, which highlights the importance of a proper parameterization of quantum solvers and quantum hardware, or \cite{salehi2021unconstrained}, stating that one of the most challenging task in quantum optimization is the choice of penalty values of the QUBO problem formulation.

When designing an experimentation in a D'WAVE quantum annealer, lot of information can be found provided by the owners of these devices about the functioning and the whole bunch of hardware-related parameters available to be tuned. Additionally, D'WAVE itself has published some reports to demonstrating the effect of the selection of several values in toy samples \cite{DWAVE}. Despite this, pieces of advice provided by the scientific community for configuring these parameters often revolve around a restricted set of them, such as the time of annealing, or the chain strength coupler in charge of keeping chains coherent. 

This is so probably due to the democratization of the quantum technologies, currently operated not only by experienced physicists but also by experts from other disciplines not familiar with the hardware specificities \cite{osaba2022systematic}. Indeed, the increasing interest on Quantum Technologies is joining efforts from scientists with different level of expertise and background. For practitioners closer to software development instead of to quantum hardware processing, it is highly recommended getting a guide of good practices when it comes to the experimentation setup and analysis. So far, good practices outlined by the community for setting some parameters are usually superficially adviced, such as \textit{the value for chain strenght should be reasonably large} \cite{willsch2022benchmarking,DWAVE}.

On this basis, and bringing to the fore the expectations placed on QC for solving real-world use cases, this work is engaged to open the frontiers of this field, previously mostly inhabited by physical experts, to non-expert stakeholders. In doing so, this analysis is driven by an experimentation focused on providing instructive responses to the requirements likely to be imposed in real settings. Such requirements are illustrated by the following premises and questions: 
\begin{itemize}[leftmargin=*]
    \item In a real environment, the obtaining of a correct and fully usable solution could be more critical than getting the optimum. 
    \begin{itemize}
        \item \textit{What is the \textbf{likelihood of reaching feasible results}, being a feasible solution an outcome which meets all the requirements and constraints of the problem\footnote{For the problem dealt in this paper, the Traveling Salesman Problem, a feasible solution is the one that visits each and every one of the nodes once and only once, and in a sequential order}? What is, indeed, the configuration that maximizes that likelihood?} 
        \item \textit{What is the \textbf{quality of the results in terms of distance to the optimum?} What is the configuration that maximizes that quality?}
    \end{itemize}
    \item In those real environments in which the obtaining of the best solution is more appreciated.
    \begin{itemize}
        \item \textit{What are \textbf{the odds of obtaining an optimum solution}?, Which is the best parameterization to obtain good 'optimum by feasible' ratios?} 
        \item \textit{How to determine a good annealer-related and hamiltonian-related parameterization in this non-deterministic framework to maximize the likelihood of reaching the optima solution?}
    \end{itemize}
    \item For those practitioners coming from real-world applications, it is mandatory to obtain a minimum of knowledge about hardware operation in order to establish good settings in the experimentation stage.
        \begin{itemize}
        \item \textit{What is the sensitivity of the parameter tuning and what are the reasons behind that behaviour? What kind of \textbf{landscape} is yielded by different parameterizations?}
        \end{itemize}
\end{itemize}

On this rationale, this paper empirically analyses the sensitivity of the hamiltonian-related parameterization (that associated to the problem at hand) and the impact of those concepts directly affecting such hamiltonian as the chain strength (with well-known influence on the performance of the algorithm) and auto-scale. Additionally, we provide behaviour-related information crucial to understand the binomial annealer-algorithm functioning such as the resulting energy distributions.

For properly guiding this experimental paper, the well-known Traveling Salesman Problem (TSP, \cite{junger1995traveling}) has been chosen as benchmarking problem. To this end, two formulations have been considered: a state-of-the-art reference expression (giving rise to a reference QUBO, r-QUBO), and a new proposed heuristic-driven expression (generating a heuristic QUBO, h-QUBO). This second formulation has been proposed with the intention of simplifying (from a classical point of view) the resolution of the TSP. In overall, 220 different configurations have been tested over more than 3700 unitary runs and 7 million of quantum reads. Some interesting findings have been drawn from this study, focused on one single TSP instance, and centered on how deemed parameters can be selected in order to increase the probability of obtaining better optimization outcomes. Besides, even though not being the main interest of this work, we have conducted an additional study on six TSP instances of 5 to 14 node with the objective of exploring the efficiency of the h-QUBO on further TSP cases.

The rest of this paper is structured as follows: in the following Section \ref{sec:problem}, we introduce the problem statement and fundamentals. Experimental setup and results are presented in Section \ref{sec:hiper}. Section \ref{sec:exp} is devoted to the second performed experimentation. Finally, Section \ref{sec:conc} closes this document by drawing some conclusions and outlining future research lines.

\section{Problem Statement and fundamentals} \label{sec:problem}

\subsection{Programming model of D'WAVE's Quantum Annealer} \label{sec:DWAVE}

This section describes the procedure to solve optimization problems on the D'WAVE device. To this end, we introduce the Ising model and the Quadratic Unconstrained Binary Optimization (QUBO) problem, which are the programming formulations for the objective function of the this quantum annealer.

In order to solve optimization problems on D'WAVE, the first step consists on mapping the problem into a binary quadratic model: the Ising model or QUBO formulation \cite{lucas2014ising}. In doing so, the coefficients of the quadratic model need to be set, so the linear terms (biases) evaluate qubits and quadratic terms  asses the interactions between pairs of qubits (couplers). 

Consider $\mathcal{Q}$ the set of qubits available in the quantum device, where each one takes spin values $s_i\in \{-1,+1\}$, and let $\mathcal{C}$ be the set of couplers between qubits. Then the Ising programing model is given by:
\begin{equation}\label{ising}
    \mathbf{H_I} = \sum_{i} h_i s_i + \sum_{(i,j)\in \mathcal{C}} J_{ij} s_i s_j,
\end{equation}
where $h_i$ is the bias for the $i$-th qubit and $J_{ij}$ the coupling strength between the $i$-th and $j$-th qubits.

It is important to highlight that in the D'WAVE quantum annealer there is no coupling between every pairs of qubits in $\mathcal{Q}$. In other words, if we consider that the set of qubits conform a graph, with $\mathcal{Q}$ the set of nodes and $\mathcal{C}$ the set of edges, the graph $G=(\mathcal{Q}, \mathcal{C})$ is not complete. Hence, in most practical applications the graph of the formulated Ising (or QUBO) model does not match with the topology of the D'WAVE device. 

To address this issue, the graph of the problem is embedded into D'WAVE's graph following a procedure called minor-embedding \cite{boothby2020next}. In this process, some variables of the problem graph are represented by a set of physical qubits. These qubits conform a chain that must act as a single qubit and so return the same value at the end of the annealing process. Towards this end, the chain strength, i.e. the coupling weight assigned to consecutive qubits in the chain, have to be set. Finding a good value of the chain strength is crucial, because if its small, the chains will break. Otherwise, if the chain strength is to large, each chain will act like a separate entity, emulating independent variables that do not interact with each other \cite{DWAVE}.

Finally, it is worth mentioning that the relation between the Ising model and QUBO formulations relays on:
\begin{equation}
    x_i = \frac{1+s_i}{2},
\end{equation}
where $x_i\in\{0,1\}$ is the binary variable for the QUBO expression. Hence, by simply applying this linear transformation to Equation \eqref{ising}, we can change spins $(s_i)$ to binary variables $x_i$ and vice versa. Formally, the objective function for the QUBO formultion is given by:
\begin{equation}
    \mathbf{H_{Q}}=\sum_i a_i x_i + \sum_{(i,j)\in\mathcal{C}}b_{ij} x_i x_j.
\end{equation} Hereinafter, unless stated otherwise, we assume QUBO notation when it comes to define the objective function of the TSP.

\subsection{Traveling Salesman Problem Formulations}\label{sec:TSP}

In order to center this work on the experimental goals specified in Section \ref{sec:Introduction}, rather than propose a new solver, we have leveraged the TSP problem and two objective functions ($\mathbf{H_1}$ and $\mathbf{H_2}$) as optimization problem definitions.

\textbf{Reference $\mathbf{H_1}$}, introducing the basic optimization problem, used in preceding research papers such as \cite{osaba2021focusing,feld2019hybrid}, or GitHub projects\footnote{https://github.com/BOHRTECHNOLOGY/quantum\_tsp}. This state-of-the-art formulation, which give rise to the r-QUBO, is hereafter detailed:

\begin{equation}
    \mathbf{H_1} = \mathbf{H_A} + \mathbf{H_B}(D_{ui}),
\end{equation}
where
\begin{equation}
    \mathbf{H_A} = A\sum_{i=1}^{n}\left( 1-\sum_{j=1}^{n} x_{i,j} \right)^2 + A\sum_{j=1}^{n}\left( 1-\sum_{i=1}^{n} x_{i,j} \right)^2
\end{equation}
and
\begin{equation}
    \mathbf{H_B}(D_{ui}) = B \sum_{(u,i)\in E} D_{ui}\sum_{j=1}^{n}x_{u,j}x_{i,j+1}.
\end{equation}
In this formulation, $x_{i,j}$ is 1 if the node $i$ is located at position $j$, while $D_{ui}$ is the distance between nodes $u$ and $i$. Finally, $E$ is the complete set of edges that composes the TSP instance. Note that this formulation is divided in two different Hamiltonian: 
\begin{itemize}
    \item $\mathbf{H_A}$ (with the penalty term $A$) requires that each node should appear once and only once in the solution.
    \item $\mathbf{H_B}$ (with the penalty term $B$) defines the order of the customer within the tour, adding to the energy function the distances of the paths $(u,i) \in E$ that comprise the solution.
\end{itemize}

\begin{figure}[htbp]
	\centering
	\includegraphics[width=0.9\columnwidth]{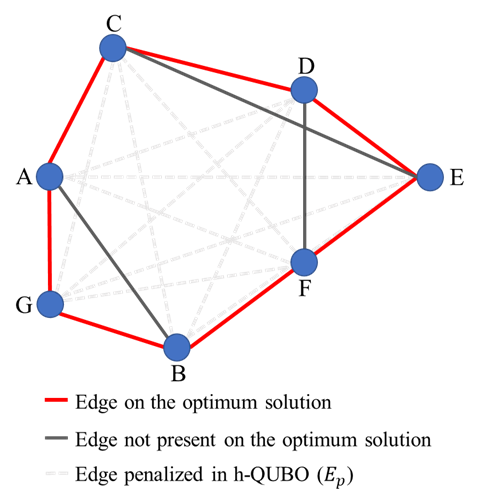}
	\caption{An overall view of the Burma'7 TSP instance considering the h-QUBO.} \label{fig:ReducedQUBO}
\end{figure}

\textbf{Heuristic-driven $\mathbf{H_2}$}, which generates h-QUBO and which tries to emulate a new reduced graph representation (such as the one depicted in Figure \ref{fig:ReducedQUBO}) through the modification of $D_{ui}$:

\begin{equation}
    \Tilde{D}_{ui} =
    \begin{cases}
     C,& \text{if } (u,i)\in E_p\\
    D_{ui},              & \text{otherwise}
    \end{cases},
\end{equation}
In contrast to the reference $\mathbf{H_B}(D_{ui})$, the component $\mathbf{H_B}$ is now a function of $\Tilde{D}_{ui}$ in charge of penalizing with parameter $C$ those unpromising connections belonging to $E_p = \left\lbrace(u,i) \mid D_{ui} > m_u\right\rbrace$, being $m_u$ the median of all the distances departing from node $u$. 

In a nutshell, it could be said that $H_A$ in both approaches delimits the feasibility of the solution, while $H_B$ and its counterpart $H_{Bp}$ is related to the optimality of the problem.

Two different benchmarking strategies have been devised for the experimentation:
\begin{itemize}[leftmargin=*]
    \item For the tests conducted to assist to the main purpose of this research (Section \ref{sec:hiper}), a single case comprised by seven nodes has been deemed (coined as \texttt{Burma'7}).
    \item For testing the h-QUBO in other scenarios (Section \ref{sec:exp}), five additional examples composed of 5, 6, 12, 13 and 14 cities have been employed.
\end{itemize}
Note that these instances have been modeled specifically for this study, adopting as base the \texttt{Burma14}, which is part of the well-known TSPLIB repository \cite{reinelt1991tsplib}. Despite being small instances in comparison to the classical computing state of the art, these amounts of nodes are coherent with current studies focused on the TSP solving in quantum computers \cite{warren2020solving,warren2021solving}. Both instances and the code employed in this research are available under demand. 

Lastly, it is interesting to point our that the amount of feasible solutions available for the case of r-QUBO is 5040. Regarding the h-QUBO, despite the number of feasible paths to the problem is still the same, just 28 non-penalized routes can be built, while 5012 solutions are feasible but penalized. This specific situation provides different landscapes in terms of paths energy. 

With the intention of increasing the understanding of the problem that we are facing, we depict in Table \ref{tab:energies} the theoretical energy for three different kind of routes: the best and worst feasible ones, and the best unfeasible solution. It should be clarified that for the case of h-QUBO, the terms feasible and unfeasible refer to non-penalized and penalized, respectively. Furthermore, the numerical value associated with each $B$ penalty has been calculated using the normalized distance matrix of \texttt{Burma'7}.

\begin{table}[ht!]
	\centering
	\caption{Energies for best and worst feasible paths, and best unfeasible solution. For h-QUBO, feasible and unfeasible refer to non-penalized and penalized, respectively.}	
	\resizebox{1.0\columnwidth}{!}{
		\begin{tabular}{c|cc|c}
			\toprule
			QUBO Type & Best feasible & Worst feasible & Best unfeasible \\
			\midrule
			r-QUBO & 2.69587$\times$$B$ - 14 $\times$$A$ & 5.58286$\times$$B$ - 14 $\times$$A$ & 2.46472$\times$$B$ - 12 $\times$$A$\\
			h-QUBO & 2.69587$\times$$B$ - 14 $\times$$A$ & 3.21168$\times$$B$ - 14 $\times$$A$ & 2.38444$\times$$B$ - 12 $\times$$A$\\
			\bottomrule
		\end{tabular}
	}
	\label{tab:energies}
\end{table}

\section{Analyzing the behaviour of the D'WAVE Quantum Annealer} \label{sec:hiper}

With the goal of empirically analyse the sensitivity/impact of the quantum-annealer to different QUBO parameter values, and inspired by studies like \cite{feld2019hybrid}, we conduct an extensive experimentation aiming to give explanatory answers to the following research questions (RQs) in the subsequent subsections:

\begin{itemize}[leftmargin=*]
    \item \textbf{RQ1}: What is the impact of problem-related parameterization in the obtaining of feasible/optimum solutions?
    \item \textbf{RQ2}: How is the difference among the user-defined QUBO and the underlying Ising model parameterization?
    \item \textbf{RQ3}: How does employing the h-QUBO impact the energy landscape?
\end{itemize}
As previously pointed, this experimentation will be done over a single instance of the TSP coined as \texttt{Burma'7}, which is composed of 7 nodes. 

Having this said, the experimentation has included linearly spaced $A$, $B$ and $\mathtt{chain\_strength}$ as problem-representation parameters, latter embedded into the QUBOs formulation:
\begin{itemize}[leftmargin=*]
    \item {\small $A \in \{0.4, 0.55, 0.7, 0.85, 1.0\}$.}
    \item {\small $B \in \{0.001, 0.001 + \Delta_B, 0.001 + 2\Delta_B, 0.001 + 3\Delta_B, 0.001 + 4\Delta_B\}$, where $\Delta_B =(A/2-0.001)/4$.}
    \item {\small $\mathtt{chain\_strength} \in \{A, A + \Delta_C , A + 2\Delta_C, A + 3\Delta_C, A+4\Delta_C\}$, where $\Delta_C = (1-A)/4$.}
    \item {\small $C = 2A/B$}, this way emulating a reduced graph by penalizing $E_p$ as much as a hard constraint (through equalising those belonging to $\mathbf{H_A}$).
\end{itemize}
As can be seen, the value of $B$ is related to the value taken by $A$ (being its maximum value $A$/2). The rationale behind this decision can be explained as follows: given that $B$ penalization is applied to the TSP problem distances, this fact results in a maximum penalized distance of $B$ (maximum normalized distance equals 1). If the quantum solver decides not to travel to a city, the optimization objective function would safe a maximum of $2B$ distance (trip to and from that city). Hence, the benefit of saving that $+2B$ should not be higher, in absolute value, that the benefit (i.e. bias) $-2A$ associated to the appearance of the variable (i.e. city) in the solution. 

The $\mathtt{chain\_strength}$ domain, in turn, starts from $A$, covering two scenarios: $\mathtt{chain\_strength} < \max(coupler)$ (i.e. $2A$) and $\mathtt{chain\_strength} \geq \max(coupler)$, in this case guaranteeing that it will be \textit{i}) as higher as the highest coupler $2A$ or \textit{ii}) the highest coupler in the Ising model itself. 

As the major interest is to get statistically significant metrics about the device outputs, we have selected a number of reads (2000) large enough to get sample significance. Furthermore, the default minor embedding procedure has applied to match the logical graph to the physical one. The time of annealing (set to 400 microseconds) and the rest of annealer-related parameters have kept their default values since their influence analysis is out of the scope of this research.

With all this, the experimentation carried out is eventually composed of more than 3700 unitary runs over the \texttt{Burma'7} instance, each run -advocated to test a specific parameter configuration- is sampled 2000 times, summing up to more than 7 million of quantum reads in overall.

As for the QC hardware, the \texttt{D'WAVE Advantage\_system6.1} computer has been used, placed in Vancouver, Canada, counting on 5627 working qubits, and being accessed by the D'WAVE Leap service\footnote{https://cloud.dwavesys.com/leap/}. 

\subsection{\textbf{RQ1}: Analyzing the effect of problem-related parameterization on the feasible/optima occurrence ratio} \label{sec:feasibility}

Feasibility odds speaks about both the landscape rendered by a specific problem-related parameterization and the solver ability to reach valid solutions, being both somehow correlated. The landscape is utterly determined by the problem definition (including parameterization) eventually encoded into a Hamiltonian. The solver ability is, on the one hand, restricted by the noise of the quantum device, and, on the other, subject to the Hamiltonian properties.

All these factors in interaction are crucial for achieving acceptable feasible ratios. This issue is challenging in quantum environments, as expressed in works such as \cite{tambunan2022quantum} or \cite{grant2021benchmarking}. This section is devoted to analyse the effect of the problem-related parameterization (involved in the hamiltonian) on the feasibility proportion.    

For the sake of brevity, and attempting to shown the most relevant information, all the analysis illustrated in these sections are restricted to those results achieved when $\mathtt{chain\_strength}> 0.8$. This narrower $\mathtt{chain\_strength}$ domain has been selected because the experimentation has empirically proved that a lower value of $\mathtt{chain\_strength}$ is directly related with poorer results. The excerpt is composed of a total of 2318 executions. 

\begin{figure}[htbp]
	\centering
	\subfigure[r-QUBO (3D surface)]{\includegraphics[width=43mm]{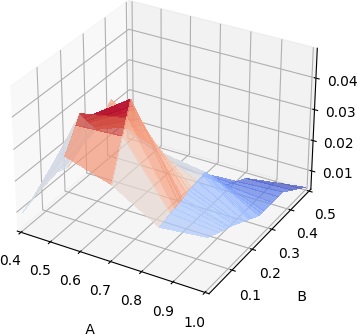}}
	\subfigure[r-QUBO (Projection)]{\includegraphics[width=43mm]{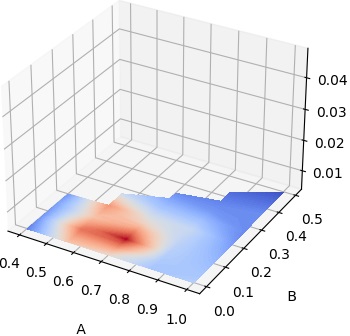}}
	\subfigure[h-QUBO (3D graphic)]{\includegraphics[width=43mm]{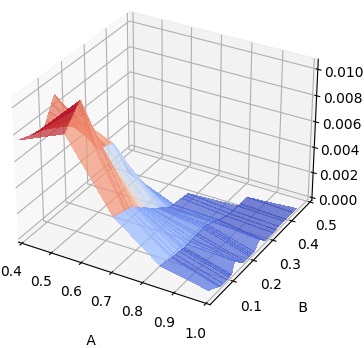}}
	\subfigure[h-QUBO (Projection)]{\includegraphics[width=43mm]{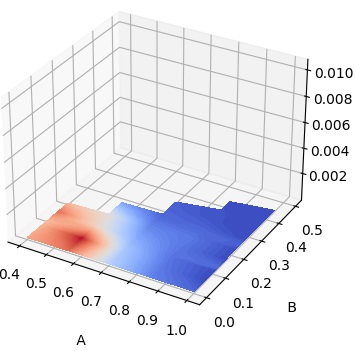}}
	\caption{Feasible ratio surface (left) and its projection (right) for $\mathtt{chain\_strength}>0.8$ and different configurations of $A$, and $B$. Results for r-QUBO are shown at the top (a and b), and for h-QUBO at the bottom (c and d).} \label{fig:feasible}
\end{figure}

Figure \ref{fig:feasible} presents the ratio of feasible solutions obtained by h-QUBO and r-QUBO. More precisely, we show in dark red the configurations with highest frequency, and in blue those with the lowest ones. Based on these results, the following conclusions are drawn:
\begin{itemize}[leftmargin=*]
    \item By observing the landscapes and projections, it becomes noticeable that the correct choosing of $A$, and $B$ is critical for maximizing the probability of finding feasible solutions. It is worth noting that depending on the configuration, the solver failed to produce any feasible result. This is specially observed for h-QUBO formulation.
    \item Furthermore, the likelihood of obtaining feasible solutions when resorting to the h-QUBO is much lower that with the r-QUBO. More concretely, at their respectively highest points, h-QUBO computation reaches 20 feasible solutions for each 2000 quantum reads (which is the number of reads employed in this study for each run), whereas this ratio grows up to 80 out of 2000 reads in the case of the r-QUBO. 
\end{itemize}
These results illustrate: i) the sensitivity of the annealer to an appropriate parameterization -- as stated in works such as \cite{mehta2019quantum} or \cite{salehi2021unconstrained} --, and ii) that r-QUBO is more prone to obtain feasible solutions when $A$ and $B$ parameters are properly set.

\begin{figure}[htbp]
	\centering
	\subfigure[r-QUBO (3D graphic)]{\includegraphics[width=43mm]{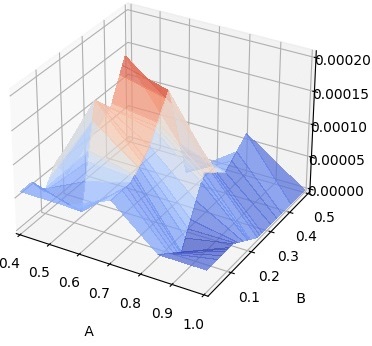}}
	\subfigure[r-QUBO (Projection)]{\includegraphics[width=43mm]{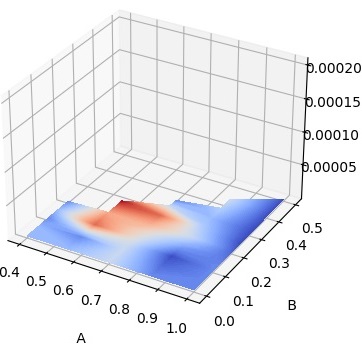}}
	\subfigure[h-QUBO (3D graphic)]{\includegraphics[width=43mm]{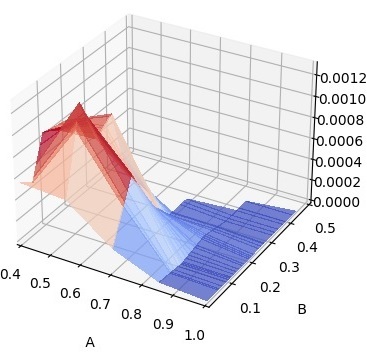}}
	\subfigure[h-QUBO (Projection)]{\includegraphics[width=43mm]{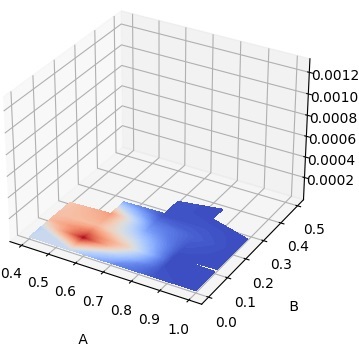}}
	\caption{Optimum ratio surface (left) and its projection (right) for different configurations of r-QUBO (a and b) and h-QUBO (b y c). The depicted results correspond to $\mathtt{chain\_strength}>0.8$.} \label{fig:optima}
\end{figure}

Turning our attention to the optimality analysis, Figure \ref{fig:optima} depicts the ratio of optimum solutions obtained for both r-QUBO and h-QUBO. Again, dark red color displays configurations with highest occurence probabilities, while dark blue depicts those with the lowest. Two important conclusions can be drawn from this preliminary analysis:
\begin{itemize}
    \item Both landscapes show increasing tendencies towards well-delimited-locations (i.e. $A$ and $B$ configurations), which gain demonstrates the sensitivity of the fine-tunning when attempting to reach a good performance regarding optimality. 
    \item The h-QUBO deployed in the quantum computer has reached 6 times more optima than the r-QUBO for their respective best $A$ and $B$ parameters. Specifically, $2.4/2000$ reads for the former and $0.4/2000$ reads for the latter. This behavior is specially noteworthy, since h-QUBO is markedly inferior in reference to feasibility.
\end{itemize}

\subsection{\textbf{RQ2}: From user-defined QUBO to Ising parameterization}\label{sec:real_parameters}

In order to fairly analyse the performance of the parameterization and latter delve into the annealer behaviour, it is important to highlight that such configuration is subject to constraints imposed by the device regarding the upper- and lower bounds of both biases and couplers (including the $\mathtt{chain\_strength}$). Namely, D'WAVE applies the following restrictions: $h  \in [-2.0, 2.0]$ and $J \in [-1.0, 1.0]$. In case that the implemented Ising model (see Equation \eqref{ising}) does not meet these requirements, \texttt{auto-scale} procedure is executed to fit the values into those allowable ranges\footnote{It is important to emphasise that D'WAVE quantum-annealer solves the Hamiltonian represented by the Ising model introduced in Section \ref{sec:DWAVE}, although we employ the QUBO formulation to map the optimization problem.}. Besides that, unless the user actively deactivates the \texttt{auto-scale} feature, this method is run for scaling all the weights even if they fit the imposed requirements. In our experimentation, we have deactivated the \texttt{auto-scale}, avoiding the expansion of the $A$, $B$ and $\mathtt{chain\_strength}$ if their values do not exceeded the allowable ranges. From this section on, the analysis is based on the rescaled $A$, $B$ and $\mathtt{chain\_strength}$ values, hereinafter referred as to $A_{\text{real}}$, $B_{\text{real}}$ and $\mathtt{chain\_strength}_{\text{real}}$). 

\begin{figure}[htbp]
	\centering
	\subfigure[r-QUBO (3D graphic)]{\includegraphics[width=43mm]{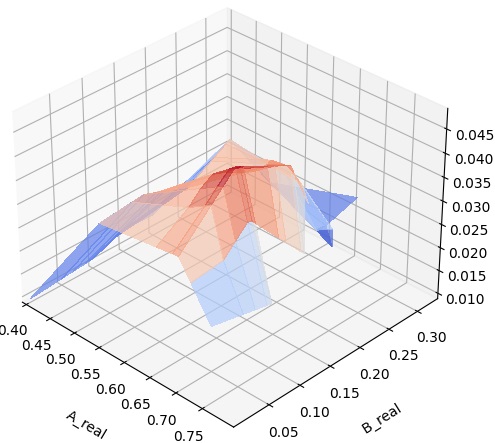}}
	\subfigure[r-QUBO (Projection)]{\includegraphics[width=43mm]{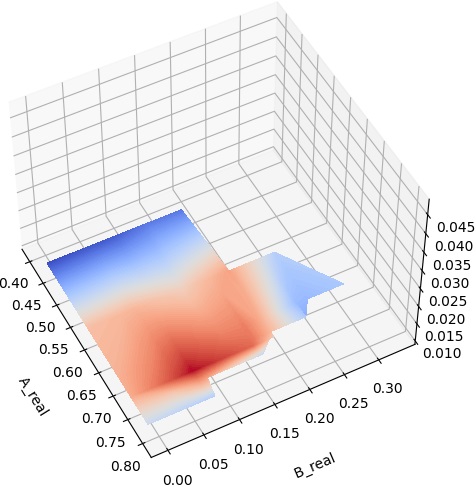}}
	\subfigure[h-QUBO (3D graphic)]{\includegraphics[width=43mm]{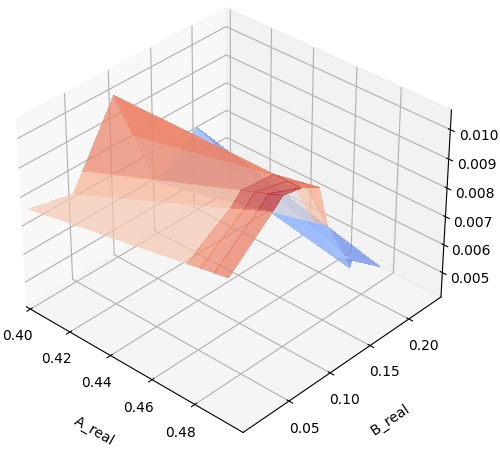}}
	\subfigure[h-QUBO (Projection)]{\includegraphics[width=43mm]{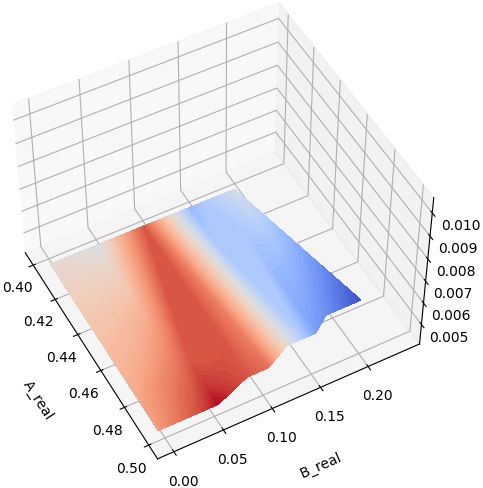}}
	\caption{Obtained feasible ratio for both QUBOs considering real AS-$A$, AS-$B$ and AS-$\mathtt{chain\_strength}$ values (considering AS-$chain\_strength>$ 0.8).} \label{fig:feasible_real}
\end{figure}

\begin{figure}[htbp]
	\centering
	\subfigure[r-QUBO (3D graphic)]{\includegraphics[width=43mm]{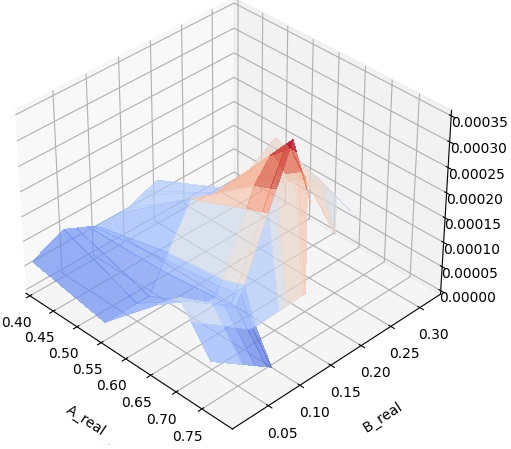}}
	\subfigure[r-QUBO (Projection)]{\includegraphics[width=43mm]{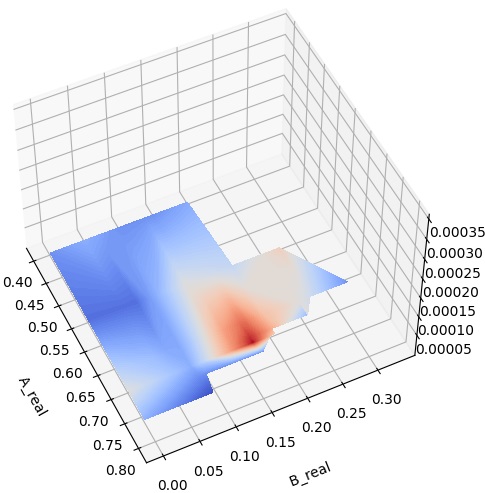}}
	\subfigure[h-QUBO (3D graphic)]{\includegraphics[width=43mm]{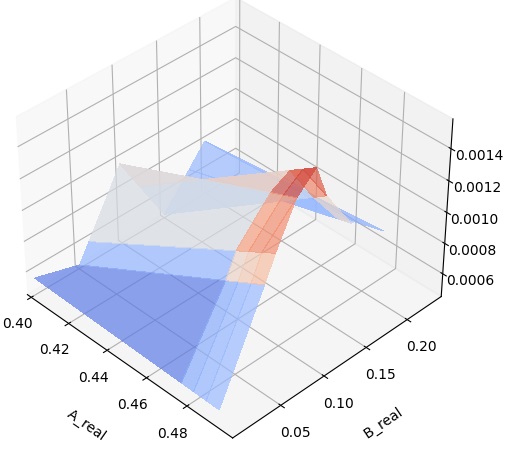}}
	\subfigure[h-QUBO (Projection)]{\includegraphics[width=43mm]{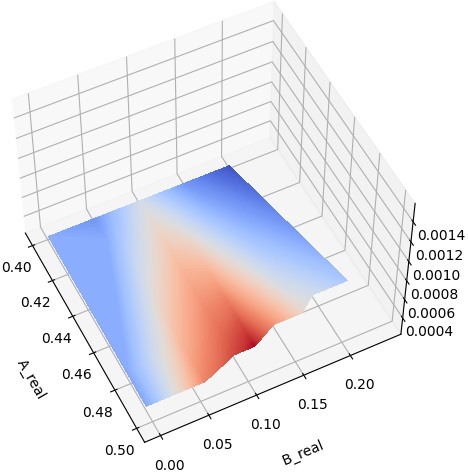}}
	\caption{Obtained optimum ratio for both QUBOs using AS-$A$, AS-$B$ and AS-$\mathtt{chain\_strength}$ values (considering AS-$chain\_strength>$ 0.8).} \label{fig:optima_real}
\end{figure}

Considering $\mathtt{chain\_strength}_{\text{real}}>0.8$, Figures \ref{fig:feasible_real} and \ref{fig:optima_real} are devoted to illustrate feasibility and optimality ratios respectively. Results depicted in Figure \ref{fig:feasible_real} support the findings reached in our previous analysis, where we show that employing the r-QUBO, the proportion of feasible solutions every 2000 reads amounts to 90, against the 20 offered by the h-QUBO. Regarding the ratios of the best found parameterizations, $A_{\text{real}}\in [0.47,0.50]$ and $B_{\text{real}}\in [0.05,0.13]$ become the best setting ranges for the h-QUBO; and $A_{\text{real}}\in [0.69,0.71]$ and $B_{\text{real}}\in [0.02,0.15]$ for the r-QUBO. Turning the attention to the optimality analysis, Figure \ref{fig:optima_real} shows that, after the parameter adjustment, conclusions with respect to optimality-in-figures remain mostly unchanged but slightly modified in landscape rendering. Optimum solutions ratio employing the h-QUBO is $2.8$ out of $2000$ reads, against the $0.7$ reached through the r-QUBO.

Finally, of particular interest is the difference between the original values of $A$ and $B$ and the scaled ones. Notice how the range of $A_{\text{real}}$ and $B_{\text{real}}$ have shrunk, in particular for h-QUBO, where the penalization in the distances $\tilde{D}_{ui}$ makes it more likely to be rescaled. Indeed, the landsdcape has swerved in their orography, currently showing different $A_{\text{real}}$ and $B_{\text{real}}$ participation in the optima reachability. In this case, the highest ratios belong to $A_{\text{real}}\in [0.47,0.50]$ and $B_{\text{real}}\in [0.10,0.13]$ for the h-QUBO; and $A_{\text{real}}\in [0.68,0.7]$ and $B_{\text{real}}\in [0.16,0.18]$ for the r-QUBO.

\subsection{A deep look into the best configurations}\label{sec:best_confs}

As preliminary conclusions about problem-related parameterization regarding both feasibility and optimality, we leveraged the five best problem-related configurations, listed in Table \ref{tab:best_confs}. For each configuration, we represent the original ($A$, $B$ and $\mathtt{chain\_strength}$) and autoscaled input values ($A_{\text{real}}$, $B_{\text{real}}$ and $\mathtt{chain\_strength}_{\text{real}}$), as well as the QUBO approach responsible for these results. Additionally, for each record, the number of optima and feasible solutions obtained is shown, along with the optimum ratio (\textit{number of optimum / number of feasible}). 

\begin{table*}[ht!]
	\centering
	\caption{Best configurations for the obtaining of optimum (top) and feasible (bottom) solutions. In this case, the Optimun ratio corresponds to  Num. Optimum / Num. Feasible.}	
	\resizebox{2.0\columnwidth}{!}{
		\begin{tabular}{ccc|ccc|cccc}
			\toprule
			\multicolumn{9}{c}{\textbf{Optimality}} \\
			$A_\text{real}$ & $B_\text{real}$ & $\mathtt{chain\_strength}_\text{real}$ & $A$ & $B$ & $\mathtt{chain\_strength}$ & Type of QUBO & Num. Optimum & Num. Feasible & Optimum ratio\\
			\midrule
			0.487 & 0.122 & 0.885 & 0.55 & 0.138 & 1.000 & h-QUBO & 53 & 324 & 0.163\\
			0.493 & 0.062 & 0.897 & 0.493 & 0.062 & 0.897 & h-QUBO & 44 & 358 & 0.122\\
			0.499 & 0.001 & 0.908 & 0.55 & 0.001 & 1.000 & h-QUBO & 38 & 393 & 0.096\\
			0.481 & 0.180 & 0.874 & 0.55 & 0.206 & 1.000 & h-QUBO & 33 & 186 & 0.177\\
			0.400 & 0.100 & 1.000 & 0.400 & 0.100 & 1.000 & h-QUBO & 33 & 335 & 0.098\\
		    \midrule
		    \multicolumn{9}{c}{\textbf{Feasibility}} \\
		    $A_\text{real}$ & $B_\text{real}$ & $\mathtt{chain\_strength}_\text{real}$ & $A$ & $B$ & $\mathtt{chain\_strength}$ & Type of QUBO & Num. Optimum & Num. Feasible & Optimum ratio\\
			\midrule
		    0.70 & 0.001 & 1.0 & 0.70 & 0.001 & 1.0 & r-QUBO & 5 & 1733 & 0.002\\
		    0.70 & 0.088 & 1.0 & 0.70 & 0.088 & 1.0 & r-QUBO & 4 & 1645 & 0.002\\
		    0.55 & 0.138 & 1.0 & 0.55 & 0.138 & 1.0 & r-QUBO & 9 & 1484 & 0.006\\
		    0.70 & 0.175 & 1.0 & 0.70 & 0.175 & 1.0 & r-QUBO & 12 & 1451 & 0.008\\
		    0.55 & 0.069 & 1.0 & 0.55 & 0.069 & 1.0 & r-QUBO & 2 & 1396 & 0.001\\
			\bottomrule
		\end{tabular}
	}
	\label{tab:best_confs}
\end{table*}

An overview on this table unveils some interesting findings coherent with those conclusions arrived at in section \ref{sec:feasibility}: the five best configurations for the optimum obtaining used the h-QUBO (supported also by the ratio between the optimum solutions and feasible solutions), whereas the r-QUBO emerges as the most appropriate alternative for getting feasible solutions. Additionally, most configurations here added, both for the original ($A$, $B$ and $\mathtt{chain\_strength}$) or the scaled ones ($A_{\text{real}}$, $B_{\text{real}}$ and $\mathtt{chain\_strength}_{\text{real}}$), fall within the most promising ranges of values spotted in Figures \ref{fig:optima}-\ref{fig:feasible_real}.

\subsection{\textbf{RQ4}: Discussion on the Energy Distribution} \label{sec:energy}

For the completeness of this study, and motivated by the conclusions in Section \ref{sec:feasibility}, we introduce this section devoted to delve into the Hamiltonian (i.e. the Ising model) properties and the subsequent annealer behaviour. To this end, we analyze the energy landscape drawn from the annealer for the two different configurations of r-QUBO and h-QUBO, and we postulate some hypotheses in order to answer those questions posed in \ref{sec:Introduction} related to: i) the likelihood of reaching feasible results, and the quality of these results. 

\begin{figure*}[ht!]
    \centering
    \centering
	\subfigure[$A_{\text{real}}=0.7$, $B_{\text{real}}=0.001$, $\mathtt{chain\_strength}_{\text{real}}=1.00$]{\includegraphics[width=0.49\linewidth]{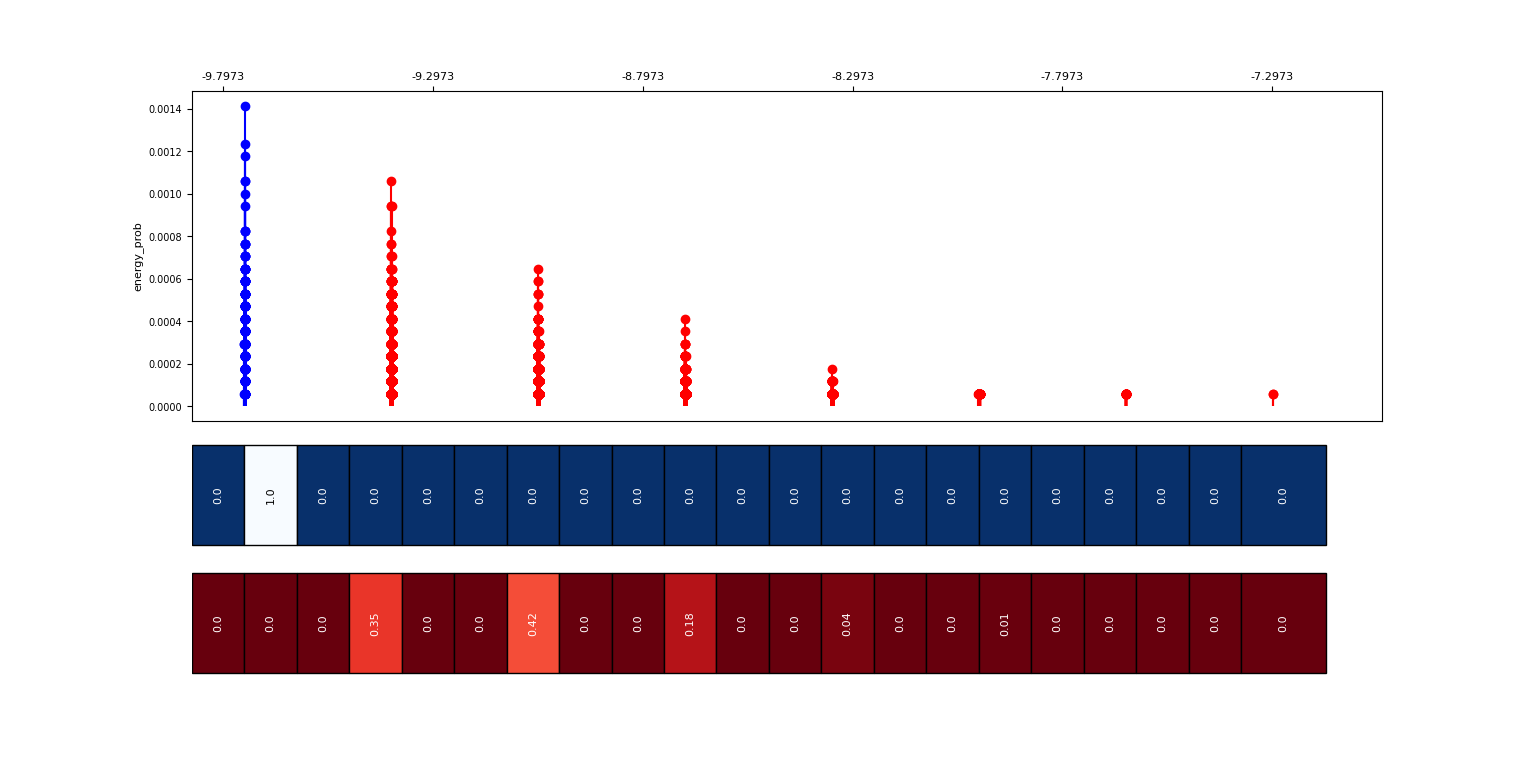}} \hfill \subfigure[$A_{\text{real}}=0.7$, $B_{\text{real}}=0.1755$, $\mathtt{chain\_strength}_{\text{real}}=1.00$]{\includegraphics[width=0.49\linewidth]{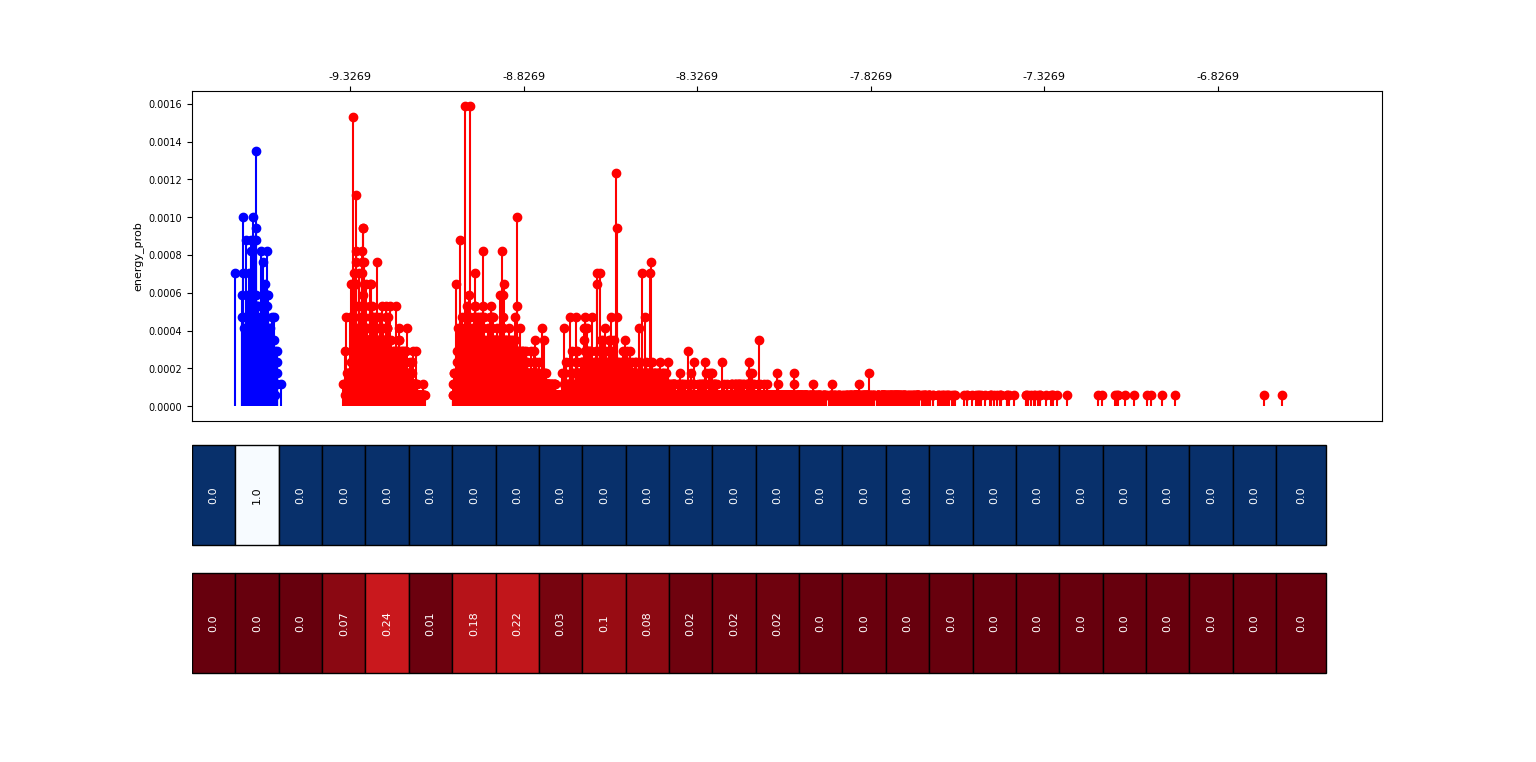}}\\
	
	\subfigure[$A_{\text{real}}=0.487$, $B_{\text{real}}=0.122$, $\mathtt{chain\_strength}_{\text{real}}=0.885$]{\includegraphics[width=0.49\linewidth]{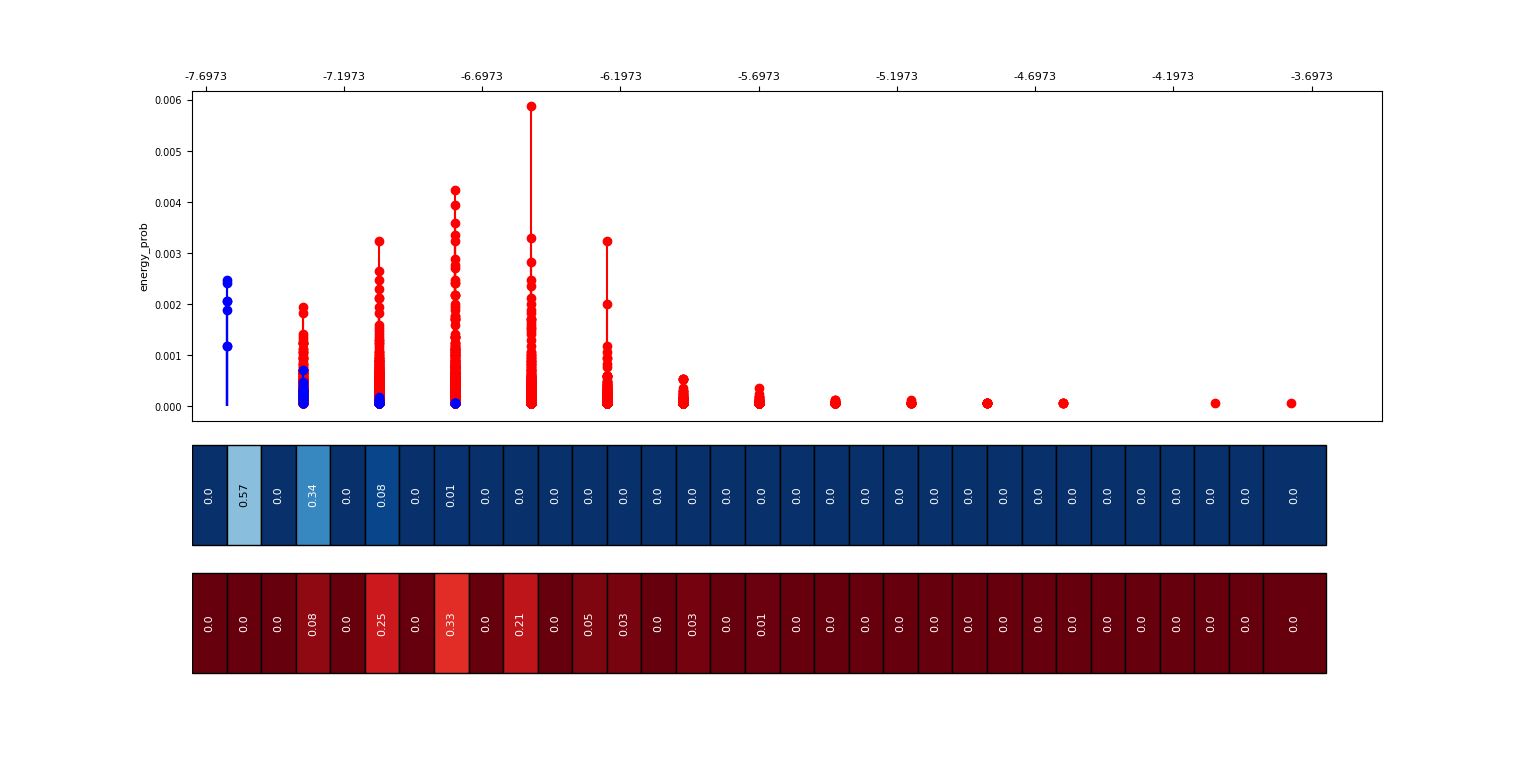}}\hfill \subfigure[$A_{\text{real}}=0.499$, $B_{\text{real}}=0.001$, $\mathtt{chain\_strength}_{\text{real}}=0.908$]{\includegraphics[width=0.49\linewidth]{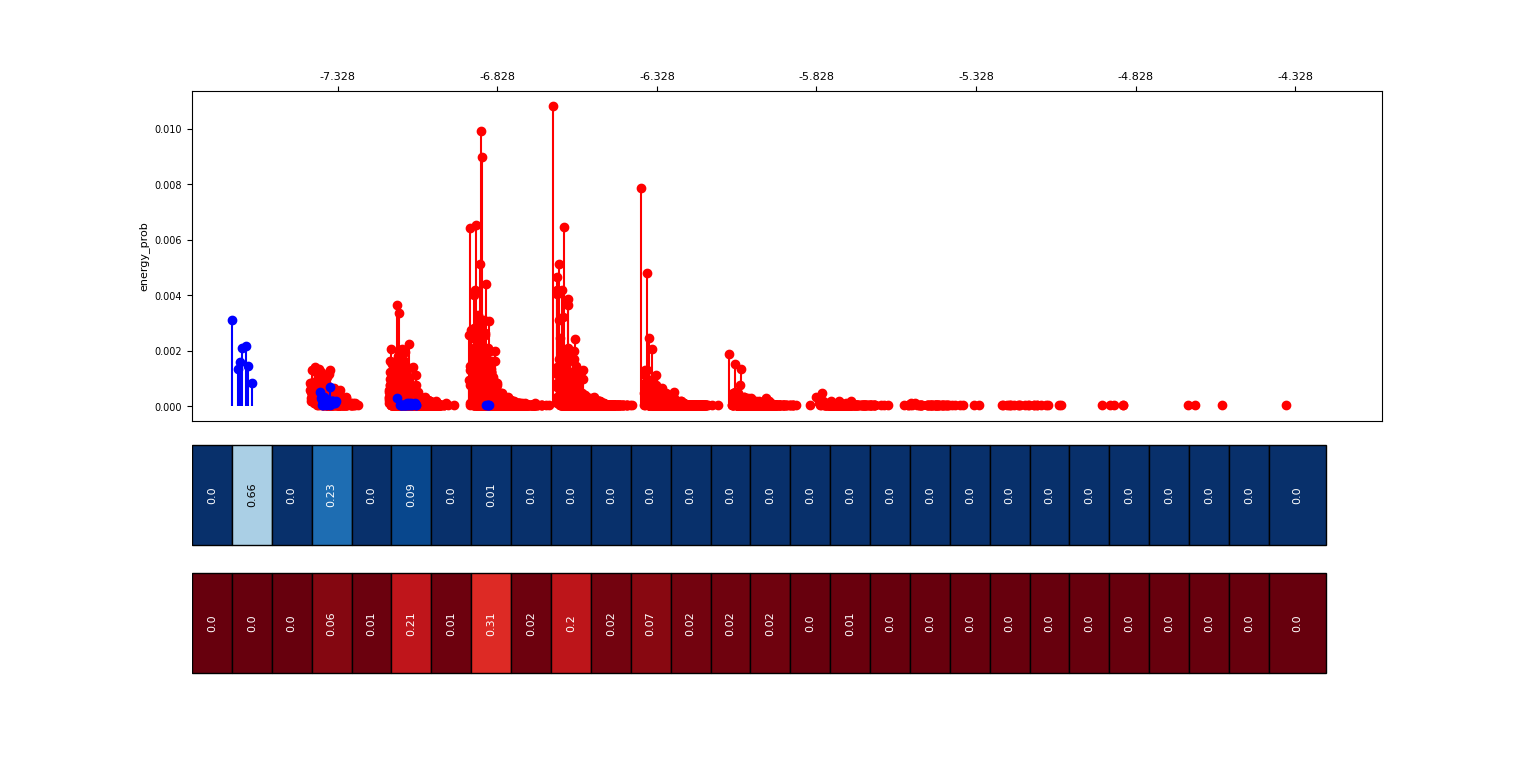}}
    \caption{Energy distribution of r-QUBO (top, a and b) and h-QUBO (bottom, c and d), for desirable configurations of feasibility (left, a and c) and optimality (right, b and d). Blue indicates feasible energy values, while red unfeasible values. In each of the figures, the first plot illustrates the occurrence frequency of experimentally obtained energies, the following plots represent the energy distribution for feasible (blue) and unfeasible solutions (red).}
    \label{fig:energy_distribution}
\end{figure*}

As we have confirmed throughout the experimentation carried out, if the objective is to maximize the probability of getting feasible solutions, in quantum environments it is highly desirable to create a wide energy gaps among feasible and unfeasible energy states.  When it comes to the D'WAVE's quantum-annealer, the minimum (energy) gap has to be, if possible, wide enough so that such distance between ground and first excited state along the process avoids jumping. Anyway, in case of jumping, it should be assured that the first excited state is another feasible one. Following these principles, we can guarantee that grouping all feasible solution together in an isolated and distant energy range conducting a fine-tuned parameterization is a good practice. This is the case for the results shown in Figure \ref{fig:energy_distribution}.a, where all feasible solutions are located into a narrow and well-delimited space. Similar results are illustrated in Figure \ref{fig:energy_distribution}.c. However, heuristically introduced penalties (in h-QUBO) make not promising solutions gain energy. Note that, for the case of h-QUBO, there are unfeasible solutions to the problem with lower energy values than those feasible not satisfying the heuristic. This fact makes the probability of obtaining feasible solutions for h-QUBO being lower than for r-QUBO.

Alternatively, if the goal is to obtain an optimum solution with high probability, the energy gap among different states is preferred to be large. The best parameter configuration for optimality results are depicted in Figure \ref{fig:energy_distribution}.b and \ref{fig:energy_distribution}.d. The former displays the results for r-QUBO, the latter for h-QUBO. In contrast to feasibility results discussed in the previous paragraph, note that the energy distribution for these configurations present a less structured landscape, with two wider energy clusters. The first one gathers feasible solutions and the second one, larger and more dense, collects more energetic states.

Finally, we evaluate now the quality of the results drawn by the quantum device by analyzing the distance to the optimum. That is, we consider that one result is better than other if its energy is closer to the optima. In this context, the higher quality results are the ones shown in Figure \ref{fig:energy_distribution}.a, as the highest occurrence probability falls within the feasible energy range. On the contrary, for configurations depicted in Figures \ref{fig:energy_distribution}.b-\ref{fig:energy_distribution}.d the most probable results are unfeasible. As a final thought, we highlight that the optimum/feasible occurrence probabilities archived in the current work has been low, even when using h-QUBO. This phenomena might be a consequence of the utilized annealing time. Further research on this line will be made in the near future.

\section{Testing the efficiency of the h-QUBO on additional TSP instances} \label{sec:exp}

As stated in the section \ref{sec:Introduction}, the main purpose of this paper is to better understand the behaviour of the D'WAVE quantum annealer when dealing with the same problem and different Hamiltonian configurations, instead of proposing a methodology for the construction of efficient QUBOs. In any case, we provide in this section a second experimentation focused on analysing the optimality when applying the heuristically-defined h-QUBO to additional instances of the TSP, in the hope of inspiring readers to heuristically built relaxed QUBOs to improve the performance of the quantum device. 

To do that, an experimentation composed of six TSP instances from 5 to 14 nodes have been executed. Each of these cases has been run 10 independent times with both approaches in order to obtain statistically representative outcomes. With respect to solver-related parameterization, the same configuration has been kept (the number of reads established in 2000 and the time of annealing set to 400 microseconds).

The problem-related parameter settings have been set after the comprehensive study conducted in the previous section to reach the best performance for each solving alternative:
\begin{itemize}[leftmargin=*]
    \item Regarding the r-QUBO, $A$ = 0.65, $B$ = 0.25 and $\mathtt{chain\_strength}$ = 1.0 have been selected.
    \item As for the h-QUBO, $A$ = 0.4 and $B$ = 0.01, while the $\mathtt{chain\_strength}$ has been set to 1.0.
\end{itemize}

Besides, as can be read in \cite{warren2020solving}, current D'WAVE computers cannot solve TSP instances with a size greater than 9 nodes \cite{jain2021solving}, because of the current qubit amount limitation. For this reason, in order to cope with \texttt{Burma'12}, \texttt{Burma'13} and \texttt{Burma'14} instances, the quantum-classical hybrid method described in \cite{osaba2021hybrid} has been adopted. This procedure relies on the partition of the whole instance into similar-sized sub-problems to be solved independently by the quantum computer. Finally, the technique builds a complete solution by merging the different sub-solutions obtained.

Results obtained in testing are depicted in Table \ref{tab:res}. For each instance and approach, we represent the global optimum as well as the average and standard deviation ($\sigma$). Moreover, to determine if the differences in outcomes are statistically significant, the Wilcoxon rank-sum test has been calculated. The obtained $Z$-value is accompanied by a single symbol: ``-'' if there is no statistical significance between the two solving alternatives and ``$\blacktriangle$'' whether the h-QUBO has yielded statistically better outcomes than the r-QUBO.

In overall, we can observe on the results that D'WAVE device reaches better outcomes when resorting to the h-QUBO. In instances greater than 6 nodes, the differences in the results are statistically significant.

\begin{table}[ht!]
	\centering
	\caption{Average and standard deviation obtained by each solving alternative, and results of the Wilcoxon rank-sum test.}	
	\resizebox{1.0\columnwidth}{!}{
		\begin{tabular}{c|c|c|c|c|c|c|c}
			\toprule
			\multicolumn{2}{c}{Instance} & \multicolumn{2}{c}{r-QUBO} & \multicolumn{2}{c}{h-QUBO} & \multicolumn{2}{c}{Wilcoxon}\\
			\cmidrule(lr){1-2} \cmidrule(lr){3-4} \cmidrule(lr){5-6} \cmidrule(lr){7-8}
			Name & Opt. & Avg. & $\sigma$ & Avg. & $\sigma$ & Res. & $Z$\\
			\midrule
			\texttt{Burma'5} & 898 & \textbf{898.0} & 0.0 & \textbf{898.0} & 0.0 & $-$ & 0.0\\
			\texttt{Burma'6} & 951 & 1061.5 & 229.9 & \textbf{970.4} & 61.3 & $-$ & -1.2603\\
			\texttt{Burma'7} & 1101 & 1519.0 & 214.92 & \textbf{1147.7} & 83.9 & $\blacktriangle$ & -3.9199\\
			\texttt{Burma'12} & 3135 & 3612.7 & 359.3 & \textbf{3172.7} & 117.6 & $\blacktriangle$ & -3.7236\\
			\texttt{Burma'13} & 3204 & 3925.8 & 477.0 & \textbf{3246.3} & 101.2 & $\blacktriangle$ & -3.7236\\
			\texttt{Burma'14} & 3294 & 4431.3 & 462.5 & \textbf{3364.9} & 153.8 & $\blacktriangle$ & -3.9199\\
			\bottomrule
		\end{tabular}
	}
	\vspace*{3mm}
	\label{tab:res}
\end{table}

\section{Conclusions and Further Work} \label{sec:conc}

\begin{table*}[ht!]
	\centering
	\caption{General overview of the experimentation carried out. }	
	\resizebox{2.0\columnwidth}{!}{
		\begin{tabular}{c|ccc|ccc}
			\toprule
			QUBO Type & $A_{\text{real}}$ range & $B_{\text{real}}$-range & $\mathtt{chain\_strength}_{\text{real}}$-range & \# real configurations tested & \# of feasible & \# of optimum \\
			\midrule
			r-QUBO & 0.4-0.7999 & 0.0007-0.3293 & 0.4-1.0 & 106 & 55655 & 223\\
			h-QUBO & 0.4-0.4999 & 0.0004-0.2375 & 0.4-1.0 & 114 & 8835 & 800\\
			\bottomrule
		\end{tabular}
	}
	\label{tab:summary}
\end{table*}

Quantum Computing is still in an incipient stage of development. Current commercial quantum computers present some important limitations, such as qubits prone to decoherence or poor error correction. Until now, the great majority of practical papers in QC are focused on addressing a specific problem or testing the adequacy of a formulation given an intuitively suitable parameterization. As a consequence, the amount of works delving on the formulation of strategies for improving the performance of QC devices is still few, even the related community has detected this topic as an interesting one.

On this experimental paper, we aim to provide some insights on the behavior of the D'WAVE quantum annealer when solving the same problem adopting different Hamiltonian configurations. To do that, we have performed a comprehensive experimentation using two different QUBOs, and focusing our attention on a single TSP instance composed of 7 nodes. For this instance, we have examined the sensitive of tuning parameters such as the penalties of the problem, and the chain strength of the QC device. 

In overall, 220 different configurations have been tested over more than 3700 unitary runs and 7 million of quantum reads. As a summary, we represent in Table \ref{tab:summary} a general overview of the experimentation carried out, divided by the type of QUBO employed.

In addition to this contribution, we have performed a further study with the goal of analyzing the efficiency of the heuristically built h-QUBO in further TSP instances. For this second experimentation, we tested the performance of the two built QUBOs through six different instances of the TSP. Obtained results have led us to conclude that using a h-QUBO, the quantum device reaches better outcomes for the problem at hand in terms of optimality.

Further research stimulated by this work is planned in several directions. Among them, we plan to adopt our proposed mechanism to additional graph-based problems, in order to deepen on the replicability of the obtained findings. Moreover, further tests will be conducted analyzing the impact of the fine-tuning of additional parameters such as the number of reads or the time of annealing. We also plan to propose alternative Hamiltonians that might improve the quality of the obtained results.

\bibliographystyle{IEEEtran}

\end{document}